# Tunable quantum dots in monolithic Fabry-Perot microcavities for high-performance single-photon sources

Jiawei Yang[1#], Yan Chen[2,3,4#], Zixuan Rao[1], Ziyang Zheng[1], Changkun Song[1], Yujie Chen[1], Kaili Xiong[2,4], Pingxing Chen[2,4,5], Chaofan Zhang[3], Wei Wu[2,4,5], Ying Yu[1,5]*, Siyuan Yu[1,5]

[1] *State Key Laboratory of Optoelectronic Materials and Technologies, School of Electronics and Information Technology, Sun Yat-Sen University, Guangzhou 510006, China*

[2] *Institute for Quantum Science and Technology, College of Science, National University of Defense Technology, Changsha 410073, China*

[3] *College of Advanced Interdisciplinary Studies, National University of Defense Technology, Changsha 410073, China*

[4] *Hunan Key Laboratory of Quantum Information Mechanism and Technology, National University of Defense Technology, Changsha 410073, Hunan, China*

[5] *Hefei National Laboratory, Hefei 230088, China*

*\*Corresponding author*: *yuying26@mail.sysu.edu.cn*

[#]These authors contributed equally to this work

## Abstract

Cavity-enhanced single quantum dots (QDs) are the main approach towards ultra-high-performance solid-state quantum light sources for scalable photonic quantum technologies. Nevertheless, harnessing the Purcell effect requires precise spectral and spatial alignment of the QDs' emission with the cavity mode, which is challenging for most cavities. Here we have successfully integrated miniaturized Fabry-Perot microcavities with a piezoelectric actuator, and demonstrated a bright single photon source derived from a deterministically coupled QD within this microcavity. Leveraging the cavity-membrane structures, we have achieved large spectral-tunability via strain tuning. On resonance, we have obtained a high Purcell factor of approximately 9. The source delivers single photons with simultaneous high extraction efficiency of 0.58, high purity of 0.956(2) and high indistinguishability of 0.922(4). Together with a small footprint, our scheme facilitates the scalable integration of indistinguishable quantum light sources on-chip, and therefore removes a major barrier to the solid-state quantum information platforms based on QDs.

## Introduction

The development of quantum light sources that are capable of deterministically producing efficient and indistinguishable photonic states is crucial for both exploring fundamental quantum physics and various applications ranging from quantum communication[1], photonic quantum computing[2] and quantum metrology[3]. Among the myriad material platforms available[4–8], semiconductor quantum dots (QDs) offer a promising way to create single-photons on-demand[9]. In addition, the manipulation of spin states in confined electrons/holes[10] or dark exciton[11] within QDs enables the creation of multi-photon entanglement, paving the way for cluster photon states[12,13]. However, as a solid-state system, QDs in intrinsic bulk material face challenges such as low photon indistinguishability[14] and low collection efficiency[15]. Addressing these issues has been a significant bottleneck in their practical utilization.

A widely adopted approach to circumvent this problem is to embed QDs into photonic cavities[15–18]. Coupling QDs, both spectrally and spatially, with the cavity mode enhances and redirects the light. Both high extraction efficiency and high indistinguishability can be achieved simultaneously for well-designed cavities. Among these architectures, open cavities and micro-pillar photonic cavities hold great promise to generate indistinguishable photons efficiently[19,20]. In an open microcavity, spectral alignment is implemented by the moving the top mirror using a nano-positioner[16]. This inevitably leads to a highly sophisticated system that is extremely sensitive to mechanical vibrations. Furthermore, scaling up these systems remains challenging due to their large footprint size. While in the micro-pillar systems, spectral overlap is typically achieved through temperature tuning[9,21,22] or the optical Stark effect[17,23], both of which are not preferred as they can degrade the quality of the photon source. Strain tuning allows for the tailoring of various QD properties in a large range without observable degradation of photon coherence or brightness[24,25]. Though attractive, the implementation of strain tuning remains challenging due to the isolated structure and high aspect ratio of micro-pillars.

Here we present a novel QD-in-microcavity structure that combines of deterministic fabrication, spectral-tunability and a high Purcell effect. We employ this structure to efficiently generate single photons. We utilize a fabrication process to integrate a positioning QD in a Fabry-Perot microcavity with a piezoelectric actuator. Small mode volume is achieved via strong lateral mode confinement facilitated by a micrometer-scale parabolic lensed-defect between two distributed Bragg reflectors (DBRs)[26]. The microcavity membrane structures allow compensation of the spectral mismatch between QD emissions and the cavity mode via a strain field, achieved by integrating QDs onto a piezo actuator. The compact and mechanically robust device makes our source less susceptible to external acoustic noise. The Purcell effect enables the generation of polarized single photons with simultaneous high extraction efficiency, high purity and photon indistinguishability. We anticipate that our unique capability to create well-ordered, tunable quantum light sources at the micrometer scale on a single substrate will significantly contribute to the advancement of scalable quantum information processing.

## Results
### Design concept.
To realize bright, tunable quantum light sources with enhanced emission rates via the Purcell effect, we have developed a new nanostructure, a monolithic Fabry-Perot microcavity, with its advantage of optimal exploitation of the Purcell effect, a compact footprint and adeptness for integration. As schematically depicted in **Fig. 1a**, the structure comprises a parabolic lensed-defect within a Fabry-Perot distributed Bragg reflectors (DBR) cavity, positioned atop a (100)-cut $[Pb(Mg_{1/3}Nb_{2/3})O_3]_{0.72}[PbTiO_3]_{0.28}$ (PMN-PT) piezoelectric actuator. In this configuration, the microcavity membrane boasts a flat morphology and a thickness of approximately ~6 μm, facilitating efficient strain transfer. In the microcavity, photons are vertically confined by top and bottom DBRs, while the lateral confinement arises from the parabolic lensed-defect in the central spacer layer (as illustrated in the spatial distribution of $E_r$ field in **Fig. 1b**). This configuration allows for a theoretically achievable high quality factor (Q-factor) of up to $10^5$ (see details in Supplementary Information Section 1, Figure S1).

The concept of a Gaussian-shaped defect Fabry-Perot cavity was first introduced in 2013[26]. Recent efforts have been devoted to create such a cavity within the GaAs material system for the utilization of QDs as single photon sources[27]. However, as a result of various growth rates along the unique crystal axis, the epitaxial layers of the top mirror become flattened after ~3 pairs-layer growth and the final cavities feature an elliptical shape. In our approach, we employ dielectric layers as the top DBR to circumvent the issue mentioned above. We can precisely engineer the cavity resonance by adjusting both the shape of the defect and the cavity's thickness (for more information, refer to Supplementary Information Section 1 and Figure S1). Moreover, the fundamental mode of the cavity exhibits a Gaussian-like far-field pattern (as seen in the inset of **Fig. 1c**), similar with micropillars, and thus enables efficient single photon extraction into free-space or single-mode fibers. For the QDs positioned in the center of the cavity, we can attain an extraction efficiency $\eta_e$ as high as 94.9% together with a Purcell factor of 40, using a 7-pair top DBR and a 46-pair bottom DBR, as shown in **Fig. 1c** (see details in Supplementary Information Section 1). With such a high factor, we can determine a near unity mode coupling efficiency $\beta$ ~0.975, using the formula $\beta=F_p/(F_p+1)$, where $F_P$ represents the Purcell factor.

**Device fabrication and characterization.**

To achieve tunable quantum light sources, we employ a transfer-printing technique. Arrays of microcavity membranes with a size of 280 μm × 280 μm are fabricated by substrate removal and are subsequently transferred onto a PMN-PT substrate using a rubber stamp (polydimethylsiloxane (PDMS)), as shown in **Fig. 2a**. Epoxy-based photoresist (SU-8) is used for the membrane bonding on the target substrate. The wide-field photoluminescence imaging technique is used to spatially overlap QDs to the cavities[28]. The positions of pre-selected single QDs are determined with respect to alignment marks using photoluminescence imaging. As depicted in the inset of **Fig. 2b**, the fluorescence image of the QDs reveals bright spots that are distinctly visible at the center of the fabricated microcavity.

We have developed a heat reflow process to prepare the parabolic lensed-defect. The photoresist is exposed around the target QDs using electron beam lithography (EBL). Reflowing at 160°C for 5 minutes renders the photoresist from disks to truncated spheres, whose topography are transferred to the $SiO_2$ layer after inductively coupled plasma (ICP) etching. We achieve precise control over the etching selectivity between the photoresist and $SiO_2$ by employing different etching chemistries. This allows the fabrication of desired aspect ratio of $SiO_2$ defects. Detailed fabrication processes are provided in Supplementary Information Section 2.

**Fig. 2c** shows a cross-section scanning electron microscope (SEM) image of a typical microcavity that containing a parabolic lensed-defect. In contract to epitaxial GaAs/AlGaAs layers, dielectric layers maintain the curved profile faithfully even after a thickness of few micrometers. The high quality of the defect can be further evidenced by atomic force microscopy (AFM) measurements in **Fig. 2d**. A very smooth surface with a roughness of only ~1.3 nm is extracted.

To experimentally evaluate the influence of defect size on the optical quality factor, we fabricate defects with various base width for a systematic study. All the samples feature a 12-pair layers in top mirror and are excited with a 785 nm continue-wave laser. The measured Q-factor increases with B and reaches a plateau at B=3.8 μm (**Fig. 2e**). A maximum Q-factor of ~15,000 was measured. Error bars represent standard deviation among various cavities. It's worth noting that in an open cavity, a maximum Q-factor of $3.1 \times 10^4$ ($\sim 6.6 \times 10^6$) is reported without (with) passivation of the GaAs surface[29]. The observed gap can be mainly attributed to the presence of non-ideal dielectric layers produced by evaporation. Further improvements can be achieved through the use of enhanced passivation methods, such as atomic layer deposition, and the deposition of high-quality films.

**Strain tunability.**

Strain is a promising method to tune QD properties by integrating QDs onto piezoelectric actuators. However, this approach is challenging to implement in isolated structures such as micro-pillars[30] or bullseye cavities[31] due to the limited transfer of strain. The microcavity structures we have developed here are compatible with strain tuning. In experiment, the strain tuning behavior is investigated by applying a voltage to the piezoelectric substrate while recording the QD emission. **Fig. 3a** shows a PL mapping of the strain-induced shift for a single QD emission line in the vicinity of the cavity resonance. Both cavity modes and the emission peak are clearly identified. Their intensity is plotted on a logarithmic scale for better viewing. Without tuning, the QD and the cavity are spectrally separated. When sweeping the voltage from -500 to 500 V, the emission peak also changes. Notably, when on resonance, nearly 50x enhancement is observed, as shown in **Figs. 3b**.

Regarding the QD's emission, a tuning range of ~1.3 nm is achieved, which can be extended more by applying large voltages. Notably, the cavity modes remain relatively constant throughout this process, which is a distinctive feature compared to temperature tuning. Possible reasons for this could include the fact that the strain is estimated to be less than 0.05%, and this level of strain does not significantly alter the morphology or the dielectric constant of the cavity.

**Single photon emission in QD coupled to monolithic Fabry-Perot microcavity.**

To develop a highly-bright polarized single-photon source, we employ the charged exciton $X^+$ and couple it to a geometrically birefringent monolithic Fabry-Perot microcavity in the Purcell regime. The mode splitting observed in the cavity is mostly induced by some residual asymmetric uniaxial strain in the semiconductor materials and elliptical geometry introduced by fabrication, a phenomenon also reported in open cavities[32]. It worth to note that this splitting can be harnessed to overcome the 50% loss limit in resonant excitation when employing cross-polarization laser filtering schemes[16,21].

**Fig. 4a** shows the photoluminescence (PL) of our microcavity modes with 7-pairs top DBR under non-resonant excitation at high-power, which splits into two modes, horizontally- and vertically-polarized (H- and V-polarized), with a separation of ~55 GHz. The linewidths of the H and V modes are $\delta\omega_H$=40.7 GHz and $\delta\omega_V$=44.5 GHz, resulting in corresponding quality factors of 8054 and 7368, respectively. Theoretically, spontaneous radiation rate of the exciton in this birefringent cavity is expected to be redistributed into H and V polarizations. On resonance, the mode is faster than the off-resonance one, resulting in a boosting a boosting factor of $1 + 4(\Delta\omega/\delta\omega)^2$~7.11[21]. Hence, a high degree polarized spontaneous emission of ~0.88 is predicted when we bring QD into resonance with the cavity H mode. The H and V polarization components of the QD's emission are shown in inset of **Fig. 4a**.

To estimate the brightness, we conduct pulsed resonance fluorescence for $X^+$ under laser excitation at an 80.1 MHz repetition rate, with the assistance of a weak 785 nm continue-wave laser to achieve nearly blinking-free single photon emission. **Fig. 4b** shows the detected single photon flux as a function of the driving laser power, revealing a complete Rabi oscillation curve. For this device, about 2.8 MHz is recorded by an avalanche photon detector (APD), which we denote as the "π pulse" condition. Considering the set-up efficiency (0.069), and avalanche photodiode correction factor (1.08) into account, an extraction efficiency of 0.58 is extracted. Detailed calibration of the system efficiency is described Supplementary Information Section 3. Residual lasers are removed by a cross-polarization setup and a long pass filter.

Furthermore, the radiative lifetime for the QD on resonance is shorten to $\tau_{on}$~100 $ps$ (blue line in

**Fig. 4c**). Compare with the average lifetime for the QDs in the slab from the same area (red line in **Fig. 4c**), a Purcell factor of ~9 is achieved. As a result, the experiment probability of emission into the H-polarized mode $\beta_H$, can be determined to be $\beta_H = F_p/(F_p + 1) \times 0.88 = 0.792$. The degree of photon polarization can be further boosted by increasing the cavity mode splitting, which can be achieved by using a defect microcavity with ellipticity (see Supplementary Information Section 1, Figure S2) or by depositing additional top DBRs to increase the Q factor of the cavity.

To further assess the purity and indistinguishability of our single photon source, the collected photons are direct to a fiber-based Hanbury Brown and Twiss setup. The second-order autocorrelation in **Fig. 4d** reveals $g^{(2)}(0)=0.044(2)$ at zero-time delay, indicating clear photon antibunching and a high degree of single-photon purity. The non-vanishing peak at zero-time delay is originated from a small amount of laser light leaking in to the detection channels, as well as re-excitation events. Furthermore, the coherence of the single photons is measured using a Hong-Ou-Mandel (HOM) interferometer, with the time separation of 12.48 ns between the two emitted single photons, which is consistent with that of the laser pulses. **Fig. 4e** shows the photon correlation histograms of normalized two-photon counts for orthogonal and parallel polarizations, indicating a raw HOM visibility of 0.811(3). Taking into account the imperfect single-photon purity and an unbalanced (52:48) beam splitting ratio in the optical setup, we calculate a corrected photon indistinguishability of 0.922(4) (see Supplementary Information Section 3). This demonstrates that generated single photons based QD in microcavity are highly coherent.

**Discussion and conclusion.**

In summary, we have developed a monolithic Fabry-Perot microcavity structure with the advantage of optimal exploitation of the Purcell effect, a small footprint and integration capabilities. By deterministic embedding of a single QD into the microcavity, we have achieved high-performance single photon sources with simultaneous high extraction efficiency, high purity, and high indistinguishability. Moreover, the far field of our microcavity exhibits Gaussian and convergent characteristics, rendering our devices compatible and advantageous for integration with fiber networks. A feasible approach could involve directly attaching the device to a fiber facet[33].

Considering future works, firstly, owing to the microcavity membrane structures, our device is naturally compatible with electrical contacts. Charge stabilization or spin injection using electrical gated-devices can be directly implemented in our devices to realize low-noise single-photon emission[34] or spin-photon entanglement/ a linear cluster state. As discussed in Ref.[13], it is anticipated that an entanglement length as long as 55-photons can be achieved by embedding the sources into microcavities with a feasible Purcell factor of 10. Secondly, strain tuning can also be employed to erase the spectral inhomogeneity between different QDs as well as eliminate the FSS, which are key elements in the realization of high-performance source of entangled photon pairs. To achieve a linewidth of the cavity that is sufficiently large to permit the extraction of both photons (biexciton and exciton), the device should be designed to have a Q factor in the range 200–300, which yields a cavity bandwidth of ∼4 nm[35]. Thirdly, the operation wavelength for both QDs and photonic nanostructures can be translated to the telecom band by changing the capping layer of QDs and the thickness of cavity and DBRs. Thus, our devices may immediately find applications in both fundamental physics and applied quantum technologies, such as quantum computing/ communication with single-photon sources, generation of spin-photon entanglement and linear cluster states for all photonic quantum repeaters[36, 37]. What is most fascinating is that the simplicity and versatility of the cavity scheme makes it possible to establish a new paradigm of manufacturing quantum light sources, in which multiple types of solid quantum light sources (including semiconductor QDs, defects et. al) with different emitter materials and operating wavelengths could be co-manufactured on the same PMN-PT platform, would potentially for future scalable quantum photonic technologies.

**Acknowledgements.**

We acknowledge Jin Liu and Yu-Ming He for the valuable discussions. We are grateful for financial support from the Science and Technology Program of Guangzhou (202103030001), the Innovation Program for Quantum Science and Technology (2021ZD0301400, 2021ZD0301605), the National Key R&D Program of Guang-dong Province (2020B0303020001), the National Natural Science Foundation of China (12074442, 12074433, 12174447), the Natural Science Foundation of Hunan Province (2021JJ20051), the science and technology innovation Program of Hunan Province (2021RC3084), and the research program of national university of defense technology (ZK21-01, 22-ZZCX-067).

**Author Contributions.**

Y. C. and Y. Y. conceived the project; J. W. Y., Y. C. and Y. Y. designed the epitaxial structure and the devices; C. K. S and Y. Y. grew the quantum dot wafers; J. W. Y. and Y. C. fabricated the devices; J. W. Y. built the optical setup; J. W. Y., Z. X. R., K. L. X., W. W. and Z. Y. Z. performed the optical measurements; J. W. Y., Y. C., C. F. Z. and Y. Y. analyzed the data; Y. Y., Y. C. and J. W. Y. prepared the main manuscript with inputs from all authors; Y. Y., Y. C., Y. J. C., P. X. C. and S. Y. Y. supervised the project.

**Conflict of Interest.**

The authors declare no conflict of interest.

# Reference.

1. Yin, J. *et al.* Satellite-based entanglement distribution over 1200 kilometers. *Science.* **356**, 1140–1144 (2017).
2. Wang, H. *et al.* Boson Sampling with 20 Input Photons and a 60-Mode Interferometer in a $10^{14}$-Dimensional Hilbert Space. *Phys. Rev. Lett.* **123**, 250503 (2019)
3. Peniakov, G. *et al.* Towards supersensitive optical phase measurement using a deterministic source of entangled multiphoton states. *Phys. Rev. B* **101**, 245406 (2020).
4. Bradac, C., Gao, W., Forneris, J., Trusheim, M. E. & Aharonovich, I. Quantum nanophotonics with group IV defects in diamond. *Nat. Commun.* **10**, 5625 (2019).
5. Santori, C., Fattal, D., Vučković, J., Solomon, G. S. & Yamamoto, Y. Indistinguishable Photons from a Single-Photon Device. *Nature* **419**, 594–597 (2002).
6. Meraner, M. *et al.* Indistinguishable photons from a trapped-ion quantum network node. *Phys. Rev. A* **102**, 52614 (2020).
7. McKeever, J. *et al.* Deterministic Generation of Single Photons from One Atom Trapped in a Cavity. *Science.* **303**, 1992–1994 (2004).
8. He, Y. M. *et al.* Single quantum emitters in monolayer semiconductors. *Nat. Nanotechnol.* **10**, 497–502 (2015).
9. Ding, X. *et al.* On-Demand Single Photons with High Extraction Efficiency and Near-Unity Indistinguishability from a Resonantly Driven Quantum Dot in a Micropillar. *Phys. Rev. Lett.* **116**, 020401 (2016).
10. Lindner, N. H. & Rudolph, T. Proposal for pulsed On-demand sources of photonic cluster state strings. *Phys. Rev. Lett.* **103**, 113602 (2009).
11. Schwartz, I. *et al.* Deterministic coherent writing of a long-lived semiconductor spin qubit using one ultrafast optical pulse. *Phys. Rev. B.* **92**, 201201(R) (2015).
12. Schwartz, I. *et al.* Deterministic generation of a cluster state of entangled photons. *Science.* **354**, 434–437 (2016).
13. Schwartz, I. *et al.* Deterministic generation of a cluster state of entangled photons. *Nat. Nanotechnol.* **354**, 434–437 (2023).
14. Kuhlmann, A. V. *et al.* Charge noise and spin noise in a semiconductor quantum device. *Nat. Phys.* **9**, 570–575 (2013).
15. Chen, Y., Zopf, M., Keil, R., Ding, F. & Schmidt, O. G. Highly-efficient extraction of entangled photons from quantum dots using a broadband optical antenna. *Nat. Commun.* **9**, 1–7 (2018).
16. Tomm, N. *et al.* A bright and fast source of coherent single photons. *Nat. Nanotechnol.* **16**, 399–403 (2021).
17. Liu, S. et al. Dual-resonance enhanced quantum light-matter interactions in deterministically coupled quantum-dot-micropillars. *Light Sci. Appl.* **10**, 158 (2021).
18. Liu, J. *et al.* A solid-state source of strongly entangled photon pairs with high brightness and indistinguishability. *Nat. Nanotechnol.* **14**, 586–593 (2019).
19. He, Y. M. *et al.* On-demand semiconductor single-photon source with near-unity indistinguishability. *Nat. Nanotechnol.* **8**, 213-217 (2013).
20. Zhai, L. *et al.* Quantum interference of identical photons from remote GaAs quantum dots. *Nat. Nanotechnol.* **17**, 829–833 (2022).

21. Wang, H. *et al.* Towards optimal single-photon sources from polarized microcavities. *Nat. Photonics* **13**, 770–775 (2019).
22. Unsleber, S. *et al.* Highly indistinguishable on-demand resonance fluorescence photons from a deterministic quantum dot micropillar device with 74% extraction efficiency. *Opt. Express* **24**, 8539 (2016).
23. Nowak, A. K. *et al.* Deterministic and electrically tunable bright single-photon source. *Nat. Commun.* **5**, 3240 (2014).
24. Lettner, T. *et al.* Strain-Controlled Quantum Dot Fine Structure for Entangled Photon Generation at 1550 nm. *Nano Lett.* **21**, 10501–10506 (2021).
25. Chen, Y. *et al.* Wavelength-tunable entangled photons from silicon-integrated III-V quantum dots. *Nat. Commun.* **7**, 10387 (2016).
26. Ding, F., Stöferle, T., Mai, L., Knoll, A. & Mahrt, R. F. Vertical microcavities with high Q and strong lateral mode confinement. *Phys. Rev. B* **87**, 1611116(2013).
27. Engel, L. *et al.* Purcell enhanced single-photon emission from a quantum dot coupled to a truncated Gaussian microcavity. *Appl. Phys. Lett.* **122**, 043503 (2023).
28. Liu, S., Srinivasan, K. & Liu, J. Nanoscale Positioning Approaches for Integrating Single Solid-State Quantum Emitters with Photonic Nanostructures. *Laser and Photonics Reviews* **15**, 2100223 (2021).
29. Najer, D. *et al.* Suppression of Surface-Related Loss in a Gated Semiconductor Microcavity. *Phys. Rev. Appl.* **15**, 044004 (2021).
30. Moczała-Dusanowska, M. *et al.* Strain-Tunable Single-Photon Source Based on a Quantum Dot-Micropillar System. *ACS Photonics* **6**, 2025–2031 (2019).
31. Moczała-Dusanowska, M. *et al.* Strain-Tunable Single-Photon Source Based on a Circular Bragg Grating Cavity with Embedded Quantum Dots. *ACS Photonics* **7**, 3474–3480 (2020).
32. Tomm, N. *et al.* Tuning the Mode Splitting of a Semiconductor Microcavity with Uniaxial Stress. *Phys. Rev. Appl.* **15**, 054061 (2021).
33. Cadeddu, D. *et al.* A fiber-coupled quantum-dot on a photonic tip. *Appl. Phys. Lett.* **108**, 011112 (2016).
34. Zhai, L. *et al.* Low-noise GaAs quantum dots for quantum photonics. *Nat. Commun.* **11**, 4745 (2020).
35. Ginés, Laia *et al.* High Extraction Efficiency Source of Photon Pairs Based on a Quantum Dot Embedded in a Broadband Micropillar Cavity. *Phys. Rev. Lett.* **129**, 033601 (2022).
36. Buterakos, D., Barnes, E. & Economou, S. E. Deterministic generation of all-photonic quantum repeaters from solid-state emitters. *Phys. Rev. X* **7**, 041023 (2017).
37. Azuma, K., Tamaki, K. & Lo, H. K. All-photonic quantum repeaters. *Nat. Commun.* **6**, 6787 (2015).

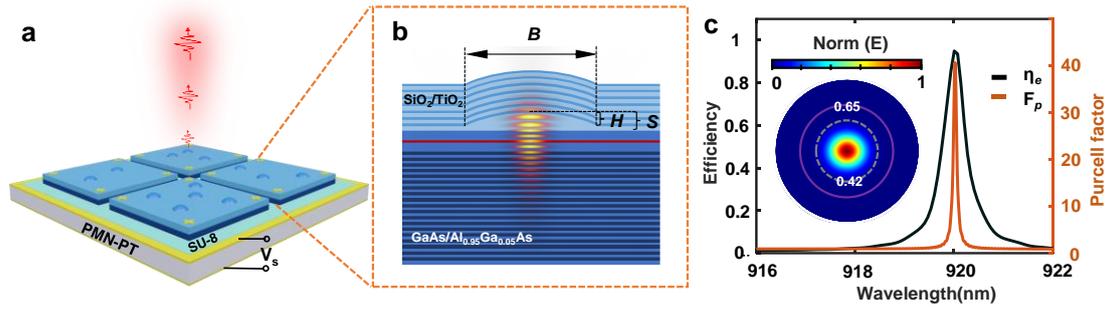

**Figure 1. The design of stain tunable single photon source. a.** Sketch of tunable single photon source. Transferable Fabry-Perot microcavity integrated with PMN-PT (100) substrate by SU-8. **b.** Cross-section of monolithic Fabry-Perot microcavity and electric field distribution of fundamental mode. The vertical confinement is from two mirrors: The top one is a dielectric $SiO_2/TiO_2$ distributed Bragg reflector (DBR), while the bottom one consists of GaAs/AlGaAs DBR. The lateral confinement is provided by the parabolic lensed-defect in the central spacer layer. Here, B is the base width of lensed defect, H is the height and S is the thickness of total $SiO_2$ spacer. The InAs/GaAs QD is positioned at the field maximum within the GaAs microcavity. **c.** Three-dimensional simulation result of the microcavity. Extraction efficiency $\eta_e$ of ~0.949 and Purcell factor $F_p$ of ~40 are obtained in device with B=4 μm, H=350 nm and S= 480nm for 7-pairs top DBR. Insect is the farfield distribution showing near Gaussian profile. Dotted gray and solid purple circles are representing NA = 0.42 and NA = 0.65, respectively.

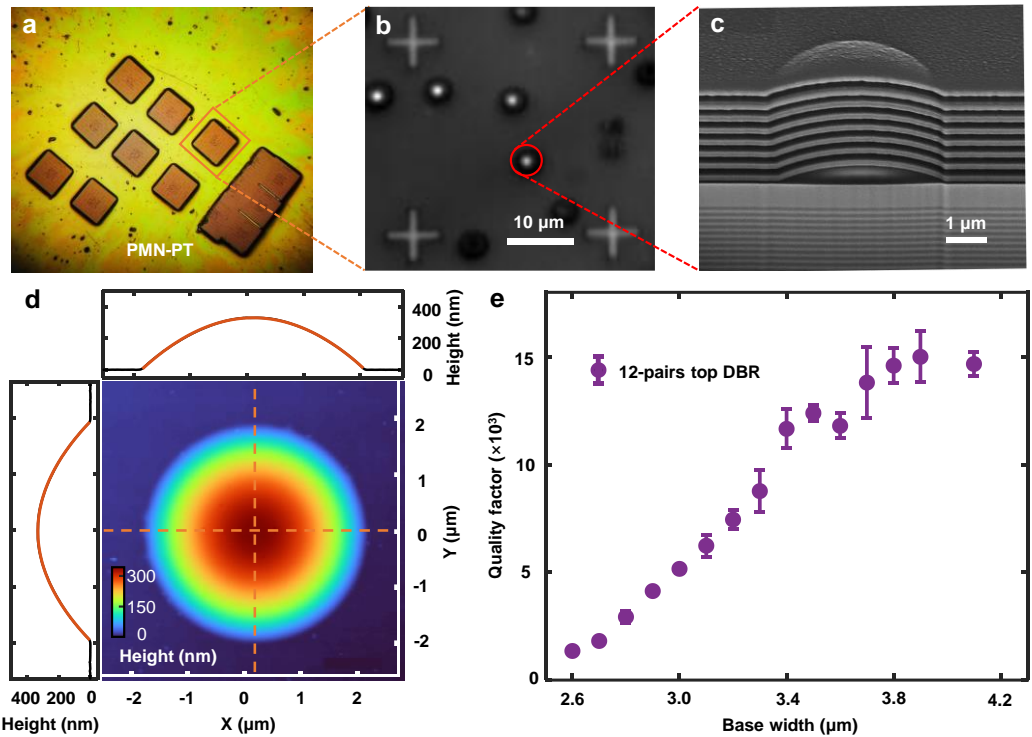

**Figure 2. Integrated demo and cavity characterization. a.** Transferable films with cavities are glued to PMN-PT (100) substrate by SU-8 adhesive. Each square film is in size of ~280 μm × 280 μm. **b.** Wide-field photoluminescence image of the fabricated Fabry-Perot microcavities with a single quantum dot in each center. The emissions from quantum dots (QDs) are excited by a high-power blue LED (445nm), while the markers are illuminated by a white LED. The adjacent markers are separated by a distance of 30 μm. **c.** Scanning microscope image of the cross section of the cavity, which is milled by focus ion beam. The lens-shaped structure is maintained even after depositing an 8-pair dielectric DBR (~2 μm thick) on top. **d.** Atomic force microscope analysis:

The color map demonstrates a near rotationally symmetric lens-shaped defect. XZ (YZ) profile is depicted in the upper (left) panel (dotted black line) and fitted to a parabola, respectively. The derived dimensions of the defect include a base width of 3.8 (3.9) μm and a height of 335 nm. One standard deviation of ~0.8 (1.3) nm of the fitted lines is extracted, revealing the paraboloid nature of the defect. **e.** Quality factor measurement: The quality factor (Q-factor) of a 12-pair $SiO_2/TiO_2$ top DBR sample is evaluated using a high-power above-band excitation with a continuous-wave laser operating at 785 nm under cryogenic conditions (4K). The mean Q-factor is plotted as a solid purple circle, with the error bar representing one standard deviation. Measurements are performed on four different devices with the same base widths. The peak Q-factor of approximately 15,000 is achieved at a base width of 3.9 μm.

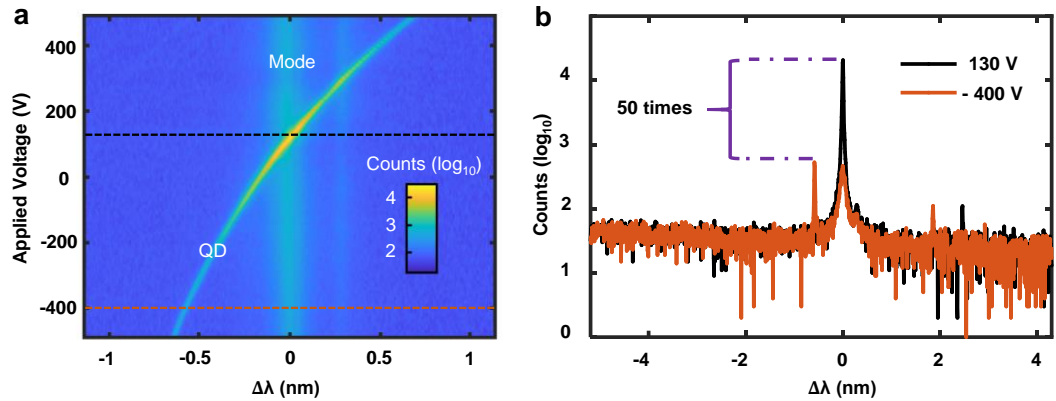

**Figure 3. Wavelength tunability using strain. a.** Stain tuning plateau: The QD emission and cavity mode are clearly identified, scaled on a log10 axis. A wavelength shift of near 1.3nm is achieved when sweeping the applied voltage from -500V to 500V. The thickness of PMN-PT (100) substrate used in this experiment is 250 μm. **b.** Spectrum extracted from (a), corresponding to the dash black line (130 V) and solid brown line (-400 V). A near 50-fold enhancement is observed when tuned into resonance.

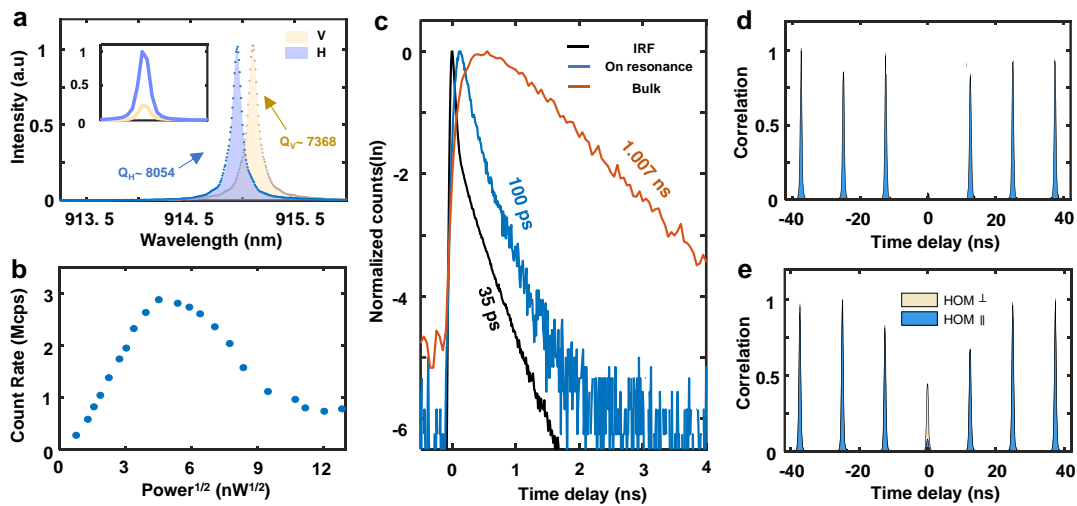

**Figure 4. Characterization of a deterministically coupled single photon source. a.** Mode splitting and QD spectrum: Under high power excitation, two linear polarized fundamental modes, horizontally (H) and vertically (V) polarized, are identified using cross-polarized techniques. The inset shows that 0.85 of the photons are directed into the H mode, while 0.15 are directed into the V mode when the QD emission ($X^+$) is in resonance with the H mode. **b.** The APD count rate of the single source is measured under pulsed resonant excitation with a repetition rate of 80.1 MHz. Rabi oscillation is observed, with the maximum population inversion

reached at 25 nW, corresponding to a π-pulse. A count rate of approximately 2.8 Mcps (million counts per second) is recorded. **c.** Time-resolved lifetime (τ) measurement: The $X^+$ lifetime histogram (blue line, τ ~100 ps) and a typical $X^+$ state lifetime curve (brown line, mean τ ~1.007 ns) in bulk material are depicted in Napierian logarithm, thus $F_p$ of ~9 is determined. Noted that seven dots in the same substrate. It is noteworthy that these measurements were conducted on seven dots within the same substrate. The instrument response function (IRF) is represented by the black line, with a τ ~5 ps. **d.** Second order autocorrelation measurement of single photons from the deterministically-coupled QD. A value of $g^{(2)}(0) = 0.044(2)$ is extracted, where the uncertainty represents one standard deviation obtained from a double-Gaussian fit to the central peak area at zero delay. **e.** HOM interference: The raw visibility of 0.811(3) is extracted from second-order correlation measurement of two consecutively emitted photons separated by ~12.48 ns.